\documentclass{aastex63}
\usepackage{bm}
\usepackage{ulem}
\usepackage{amsmath}

\usepackage{comment}

\usepackage[encapsulated]{CJK}

\shortauthors{Kino et al.}
\shorttitle{Constraining a secondary black hole mass in M87}

\received{2025 March 7}
\revised{2025 April 13}
\accepted{2025 April 18}

\begin{document}

\title{Constraining the Mass of a Hypothetical Secondary Black Hole in M87 with the NANOGrav 15-Year Data Set}

\correspondingauthor{Motoki Kino}
\email{motoki.kino@gmail.com}

\author[0000-0002-2709-7338]{Motoki Kino}
\affil{Kogakuin University of Technology \& Engineering, Academic Support Center, 
 2665-1 Nakano-machi, Hachioji, Tokyo 192-0015, Japan}
\affil{National Astronomical Observatory of Japan, 2-21-1 Osawa, Mitaka, Tokyo 181-8588, Japan}

\author[0000-0003-2938-7096]{Masahiro Nagashima}
\affil{Faculty of Education, Bunkyo University, 3337, Minami-ogishima, Koshigaya, 
Saitama 343-8511, Japan}

\author[0000-0002-7322-6436]{Hyunwook Ro} 
\affil{Korea Astronomy \& Space Science Institute, Daedeokdae-ro 776, Yuseong-gu, Daejeon 34055, Republic of Korea}

\author[0000-0001-6311-4345]{Yuzhu Cui}
\affil{Institute of Astrophysics, Central China Normal University, Wuhan 430079, China}
\affil{Research Center for Astronomical Computing, Zhejiang Lab, Hangzhou 311100, China}

\author[0000-0001-6906-772X]{Kazuhiro Hada}
\affil{Graduate School of Science, Nagoya City University, Yamanohata 1, Mizuho-cho, Mizuho-ku, Nagoya, 467-8501, Aichi, Japan}
\affil{Mizusawa VLBI Observatory, National Astronomical Observatory of Japan, 2-12 Hoshigaoka-cho, Mizusawa, Oshu, 023-0861, Iwate, Japan}

\author[0000-0001-6558-9053]{Jongho Park}
\affil{School of Space Research, Kyung Hee University, 1732, Deogyeong-daero, Giheung-gu, Yongin-si, Gyeonggi-do 17104, Republic of Korea}
\affil{Institute of Astronomy and Astrophysics, Academia Sinica, P.O. Box 23-141, Taipei 10617, Taiwan}

\begin{abstract}

Galaxy mergers, each hosting a supermassive black hole (SMBH), are thought to form SMBH binaries.
Motivated by recent observations from the East Asian VLBI Network (EAVN) showing periodic behavior in the M87 jet, a precession of about 11 years and a transverse oscillation of about 0.9 years,  we constrain the mass of a hypothetical secondary black hole orbiting the primary SMBH in M87.
To constrain the mass ratio between the primary SMBH ($M_{1}$) and the secondary black hole ($M_{2}$) defined as $q \equiv M_{2}/M_{1} \leq 1$, and the length of the semimajor axis of the binary system ($a$), we impose the following three constraints:
(i) the lower limit of $a$, below which the 
SMBH binary is expected to merge.
(ii) the strain amplitude of the gravitational wave background (GWB) at nanohertz frequencies
shown in the NANOGrav 15-year dataset.
(iii) 
a finite length of the semimajor axis of $M_{1}$, that can induce periodic behavior in the jet.
By combining these constraints, we obtain the allowed parameter space for $q$ and $a$.
If either of the EAVN-detected periods ($T$) corresponds to the binary's orbital period, 
the allowed range of $q$ is $6.9 \times 10^{-3} \le q \le 
4.2 \times 10^{-2}$ for $T \approx 11$ years, and 
$3.7\times 10^{-2} \le q \le 1$
for $T \approx 0.9$ years.
VLBI astrometric monitoring of the jet base of M87 is essential to explore the allowed parameter space for $q$ and $a$.

\end{abstract}

\keywords{
Active galactic nuclei (16);
Gravitational wave astronomy (675);
Radio interferometry (1346); 
Supermassive black holes (1663);
Very long baseline interferometry (1769)
}

\section{Introduction}
\label{sec:intro}


Growing observational evidence suggests that most massive galaxies contain supermassive black holes (SMBHs) at their centers \citep[][]{kormendy95, richstone98}. 
In the standard model of hierarchical structure formation, frequent galaxy mergers are expected
\citep[e.g.,][]{ostriker77,lacey93}, which can lead to the formation of SMBH binaries 
\citep[e.g.,][]{begelman80,milosavljevic01}. 
Multiple mechanisms are capable of shrinking the binary orbital separation through  intermediate
stage(s), solving the so-called final-parsec problem 
\citep[e.g.,][for review]{milosavljevic03}. 
Possible scenarios would be,
interaction of the binary with a gas disk 
\citep[e.g.,][]{gould00,armitage05}, 
a massive perturber
\citep[e.g.,][]{goicovic17,bonetti18}, 
and nonaxisymmetric stellar distributions that allow a high interaction rate between stars and the binary 
\citep[][]{Gualandris17}.
At the last stages of their orbital evolution, binaries produce nanohertz gravitational-wave (GW) emission, which can be detected by pulsar timing arrays (PTAs) that systematically monitor a large number of millisecond pulsars.

The NANOGrav 15-year dataset actually provides
the evidence for the presence of a low-frequency gravitational-wave background (GWB) \citep{agazie23a}.
The inferred GWB amplitude and spectrum are 
broadly
consistent with astrophysical expectations for a signal from a population of SMBH binaries, although more exotic cosmological and astrophysical sources cannot be excluded \citep{agazie23b}.
Astrophysically motivated models of SMBH binary populations can reproduce both the amplitude and shape of the observed low-frequency gravitational-wave spectrum. 
Despite strong theoretical and observational support for the pairing of SMBHs following galaxy mergers, definitive evidence for the existence of close-separation SMBH binaries approaching merger remains elusive.
Therefore, the next crucial step should be a focused search for these SMBH binaries. 
The GWB is expected to be strongly influenced by SMBH binaries at low redshift \citep[][]{wyithe03,sesana04,enoki04}. 
In the nano-Hz range,  the GWB is primarily contributed by the population of
low-redshift massive SMBH binaries
with masses greater than  $10^{9}~M_{\odot}$ \citep[e.g., see 
Figure~3 in][]{enoki04}.
Hence, systematically narrowing the allowed mass range for companion black holes in massive SMBHs at low redshift is highly significant. 
This study aims to take the first step toward this goal.
In this context, it is intriguing to investigate the presence or absence of a secondary black hole in M87, one of the most massive SMBHs in nearby galaxies \citep[][]{EHT19_1,EHT24,hada24}.
It has been reported that the luminous center of M87 and its active galactic nucleus (AGN) are offset, suggesting that the SMBH may not be located at the galaxy's center of mass \citep[][]{batcheldor10}. This displacement could be due to residual gravitational recoil oscillations following a merger event \citep[][]{lena14}.

Interestingly, recent monitoring of the M87 jet using the East Asian VLBI Network (EAVN) at 22 and 43~GHz has revealed the presence of periodic features \citep{ro23_trans, cui23}. 
The presence of periodicities is one of the most direct indicators of the orbital motion of a SMBH pair \citep[e.g.,][]{begelman80, dorazio18}, suggesting that they may be linked to the orbital dynamics of an SMBH binary system.
If M87 indeed hosts a SMBH binary system, the orbital period of the binary is expected to manifest in the periodic features 
detected in the M87 jet. 
Our ultimate aim is to identify sources of GWB detected by the NANOGrav. 
To this end, the immediate objective of the present work is to investigate the possible range of the mass ratio  $q$ and the semimajor axis $a$ for the hypothetical SMBH in M87, 
the closest and most suitable example.

In \S~2, we briefly overview the  model of binary orbital evolution following \citet{zhao24}.
In \S~3, we summarize observational constraints for  M87 that help constrain the allowed parameter space for $q$ and $a$.
In \S~4, we present the obtained constraints on $q$ 
for M87. 
In \S~5, we discuss implications from the results.
In \S~6, we summarize our findings.

\section{SMBH binary system}\label{sec:lengths}

\subsection{Basic quantities of SMBH Binary}

\begin{figure} 
\includegraphics
[width=18cm]
{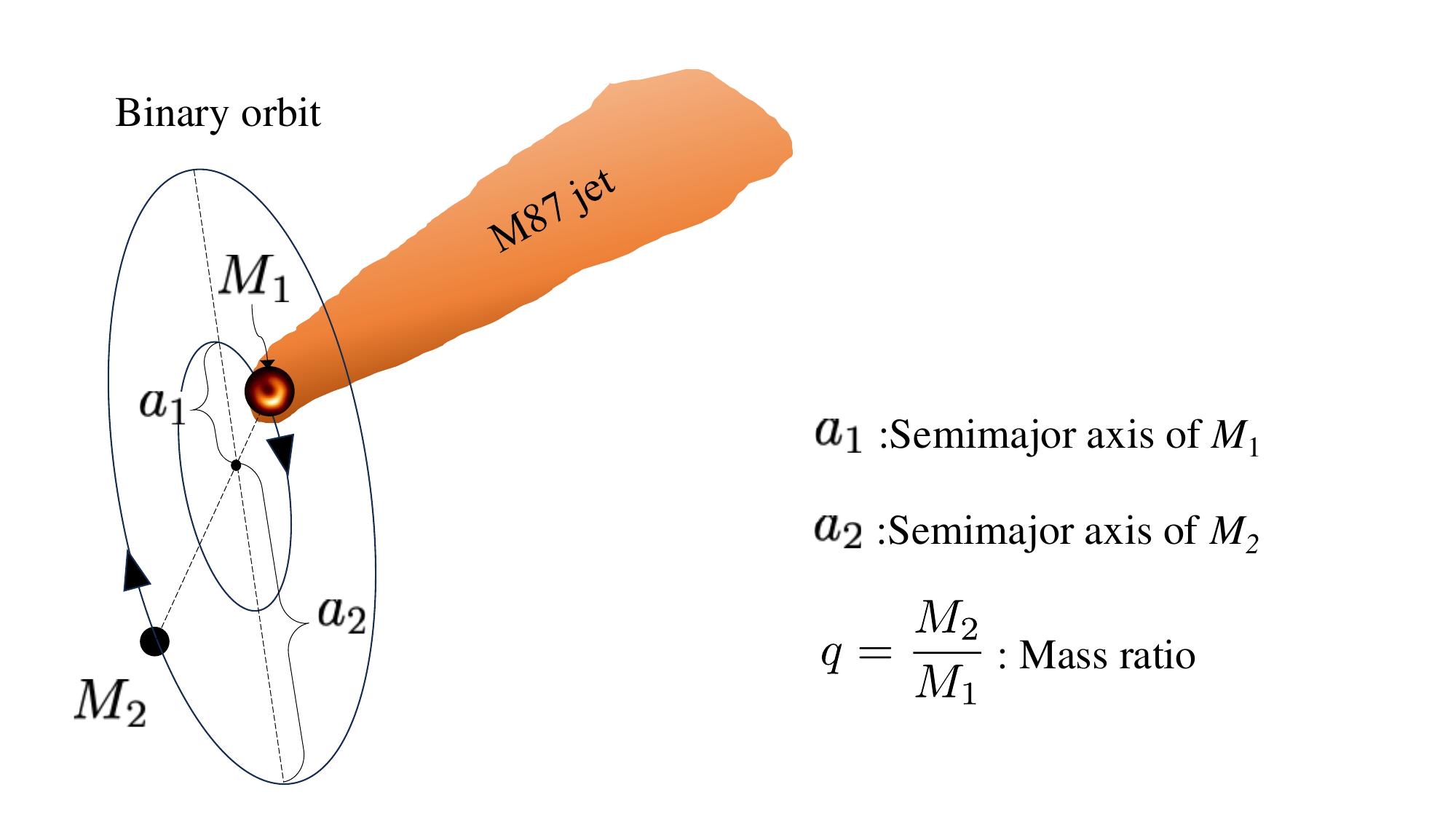}
\caption
{An illustration of the basic geometry of the hypothetical SMBH binary system considered in this study, with M87 used as a prime example. 
The primary black hole ($M_{1}$) generates the prominent radio jet observed at low frequencies, with the jet base anchored to $M_{1}$,
while the secondary black hole ($M_{2}$) does not produce a jet. 
The reflex motion of $M_{1}$ likely induces periodic behaviors in the jet, such as precessing motion, transverse oscillation, and other similar effects
\citep[e.g.,][and references therein]{ressler24}.
The M87 photo embedded in $M_{1}$ 
is adopted from \citet{EHT24}.}
\label{fig:geometry}
\end{figure}

Throughout this work, we assume a SMBH binary system 
in a circular orbit for simplicity. 
The binary's total mass ($M$),
semimajor axis ($a$),
and the mass ratio ($q$)
of the primary SMBH ($M_{1}$)
and secondary SMBH ($M_{2}$) are defined by
\begin{align}
M\equiv M_{1}+M_{2},\quad
    a \equiv a_{1}+a_{2}= \frac{1+q}{q}a_{1}, \quad
    q \equiv \frac{M_{2}}{M_{1}} \quad (0 < q \le 1)  ,
\end{align}
where $a_{1}$ and $a_{2}$ are the semimajor axis of 
the orbit of
the primary and secondary SMBH, respectively
(see Figure~\ref{fig:geometry}).
When $q=1$, $ a=2a_{1}$ holds.
The angular velocity of the circular orbital motion
($\omega$) is given by $\omega^{2}= G M/a^{3}$.
Therefore, the orbital period of the SMBH binary ($T$) 
and  the angular scale of the semimajor axis 
on the sky ($\theta$) are estimated as
\begin{align}\label{eq:T}
T=  \frac{2\pi}{\omega}= 17.6~
\left(\frac{a}{10^{17}~{\rm cm}}\right)^{3/2}
\left(\frac{M}{10^{9}~M_{\odot}}\right)^{-1/2}
~{\rm yr},
\end{align}
and
\begin{align}
\theta \equiv
\theta_{1}+\theta_{2} = 
67~\left(\frac{a}{10^{17}~{\rm cm}}\right)
\left(\frac{D}{100~{\rm Mpc}}\right)
~{\rm \mu as}  ,
\end{align}
where 
$\theta_{1}=a_{1}/D$,
$\theta_{2}=a_{2}/D$, and 
$D$ is the source distance from the Earth.
This angular scale is well achievable by 
VLBI astrometric observations  
\citep[e.g.,][for review]{rioja20}.

\subsection{Binary orbital evolution}


We provide a brief overview of
a model describing the 
orbital evolution (shrinkage) of SMBH system
following the previous works \citep[e.g.,][]{begelman80, dorazio18,zhao24}.

On kilo-parsec scales, dynamical friction is known to be the primary mechanism for angular momentum loss \citep{chandrasekhar43}.
As the separation $a$ decreases, dynamical friction 
becomes less effective, and individual interactions between each star and the binary system should be 
more effective. 
The timescale of the hardening can be given by
$t_{\rm hard}=
4\pi r_{\rm core}^{2}/(9C\sigma_{\star} a)$
where 
$r_{\rm core}$,
$C$, and 
$\sigma_{\star}$ are
the core radius,
the hardening rate coefficient, and
the stellar velocity dispersion of the host galaxy, respectively
\citep{quinlan96,zhao24}.
To be a binary system, the separation should be shorter than the hardening radius $a_{\rm hard}$ as the boundary between the dynamical friction process and the hardening state, which is given by
\begin{align}
    a_{\rm hard} = 
\frac{GM_{1}}{3\sigma_{\star}^{2}}  .
\end{align}
The binary system is formed when $a$ decreases to 
the range of $a\le a_{\rm hard}$
\citep[][]{begelman80}.

In a gas-driven case, the simple assumption is that the
binary orbit shrinks via interaction with the environment, either by gas accretion or by application of positive torque to a circumbinary disk 
\citep[e..g,][]{rafikov16}.
This process is highly uncertain 
\citep[e.g.,][]{miranda17,tang17, dorazio18}.
Following \citep[][]{zhao24}, here we set the gas-driven 
orbital decay timescale ($t_{\rm gas}$)
\begin{align}
    t_{\rm gas} 
= \frac{q}{(1+q)^{2}} \frac{1}{\dot{m}} t_{\rm Edd}, \quad
\dot{m}\equiv \frac{\dot{M}}{\dot{M}_{\rm Edd}} ,
\end{align}
where $\dot{m}$ is the mass accretion rate $(\dot{M})$ normalized
by Eddington accretion rate
$(\dot{M}_{\rm Edd})$, and 
the Eddington time,
$t_{\rm Edd}\equiv M/\dot{M}_{\rm Edd}\approx 4.5\times 10^{7}~ {\rm yr}$, 
is the time it takes for surrounding gas to accumulate until its total mass reaches the total mass of the binary system (i.e., $M$) at the Eddington accretion rate
with the accretion efficiency of $10\%$
\citep[][]{dorazio18}. 
From the relation 
of $t_{\rm hard} = t_{\rm gas}$, 
one can obtain $a_{\rm gas}=\frac{16\pi r_{\rm core}^{2}\dot{m}}{9Ct_{\rm Edd}\sigma_{\star}}\frac{(1+q)^{2}}{4q}$.
\footnote{Note that this form neglects the loss-cone depletion for simplicity, which will not affect the following arguments
\citep{zhao24}.}
When the separation decreases to $a_{\rm gas}$, there are no stars in the region and gaseous environment plays the key role in the orbital evolution.

When the separation further decreases to $a_{\rm GW}$, 
the binary system begins to lose its angular momentum
via GW emission.
The timescale for GW radiation is given by
\begin{align}\label{eq:t_GW}
t_{\rm GW} = 
\frac{5}{64} 
\frac{c^{5} a^{4}}{G^{3}M^{3}} 
\left[\frac{4q}{(1+q)^{2}}\right]^{-1}  ,
\end{align}
by \citet{peters64}.
This approximation is adequate for this work, since post-Newtonian corrections do not become appreciable until the final day scale of the merger \citep[e.g.,][]{kocsis08}.
A criterion for GWs becoming important for binary evolution can be estimated as 
$t_{\rm GW}$ becomes shorter than
the age of the Universe
$t_{\rm Univ}\approx 1.38\times 10^{10}$~yr
\citep[e.g.,][for review]{dorazio23}. 
For M87, this happens where $a\sim 10^{18}$~cm
(see Section \ref{sec:result} for details).
%

\section{Observational constraints for M87}

In preparation for applying the aforementioned SMBH binary model to M87, we summarize relevant observational constraints for M87.
Hereafter, 
we assume that that the black hole shadow observed by the Event Horizon Telescope (EHT) corresponds to the primary SMBH. 
We denote the EHT-constrained black hole mass as $M_{\rm EHT} \equiv (6.5\pm0.7) \times 10^{9}~M_{\odot}$
\citep[][]{EHT19_1,EHT19_2,EHT19_3,EHT19_4,EHT19_5,EHT19_6}
and adopt this as the mass of the primary SMBH, $M_{1} = M_{\rm EHT}$.
The gravitational radius of the primary SMBH is denoted by $r_{g}$.
This corresponds to the angular scale of gravitational radius
$\theta_{g}\equiv GM_{\rm EHT}/c^{2}D=3.8\pm 0.4~{\rm \mu as}$ at the distance of $D=16.8$~Mpc
\citep{blakeslee09,EHT19_6}.

\subsection{Periodic behaviors of the jet}

As mentioned in the introduction, the recent monitoring program of the M87 jet using EAVN at 22 and 43~GHz have detected periodic behaviors in the jet \citep{ro23_trans,cui23}.
\citet{cui23} investigated the time sequence of 170 VLBI images of the M87 jet obtained by the EAVN and other VLBI observations between 2000 and 2022. It is found that the position angle of the jet direction near the core changes with the precession period of 
\begin{equation}
T_{\rm prec} = 11.24\pm 0.47 ~{\rm years} .
\end{equation}
\citet{ro23_trans}
monitoring of the M87 jet at KVN and VERA Array (KaVA) 22 GHz from December 2013 to June 2016 with the average time interval of 0.1 year and they have found that the ridge lines show transverse oscillations with the period of 
\begin{align}
T_{\rm trans} = 0.94 \pm 0.12~{\rm years} .
\end{align}
The reflex motion of the primary SMBH $M_{1}$ could induce oscillatory motion in the radio jet.
For convenience, we define the semimajor axis $a_{T}$ corresponding to a binary orbital period $T$, given by
\begin{align}\label{eq:a_T}
a_{T} \equiv
\left(\frac{GM T^{2}}{4\pi^{2}}\right)^{1/3}
=\left(\frac{GM_{1}T^{2}}{4\pi^{2}}\right)^{1/3}
(1+q)^{1/3}   ,
\end{align}
which is essentially the same as Equation~(\ref{eq:T}).

\subsection{Tentative constraints on $a_{1}$ by VLBI astrometry}

The presence of a secondary SMBH induces reflex motion of the primary SMBH, $M_{1}$, around the center of mass with the radius of $a_{1}$
(see Figure~\ref{fig:geometry}).
Although constraining $a_{1}$ is challenging, VLBI phase-referencing observations can provide upper limits on $a_{1}$.
\citet{hada12} can provide a unique comparison of 2 epochs data. 
The core position remained stable on 
$\sim 30~\mu$as (equivalent to $\sim 8~r_{g}$) 
as on the projected scale on the sky during the 10 days.
A similar result was reported during the VHE flare in 2008 that remained stable within $\sim 45~\mu$as
 (equivalent to $\sim 12~r_{g}$) 
\citep{acciari09}.
These studies tentatively place an upper limit of $a_{1}$.
However, it should be noted that the constraints on $a_{1}$ derived from previous VLBI astrometric observations have significant limitations.
The primary limitation is the short monitoring duration, which makes it difficult to investigate binary systems with long orbital periods.
We will discuss this issue in more detail in the next section.

\subsection{The NANOGrav 15-years dataset}\label{sec:GWB}

Following \citet{schutz16}, one can place constraints on a secondary black hole using the GWB strain amplitude obtained by PTA observations.
The amplitude of continuous GWs can be parameterized by the dimensionless characteristic strain amplitude $h_{c}$, averaged over 
the whole sky \citep[e.g.,][]{jenet06}.
For a SMBH binary at leading post-Newtonian order under the assumption of circular orbits and evolution purely by energy loss via GW radiation, the strain amplitude is given by
\begin{align}\label{eq:hc}
h_{c} &= 2.76\times 10^{-14}
\left(\frac{{\cal M}_{\rm ch}}{10^{9}~M_{\odot}}\right)^{5/3}
\left(\frac{D}{10~{\rm Mpc}}\right)^{-1}
\left(\frac{f_{\rm GW}}{10^{-8}~{\rm Hz}}\right)^{2/3}, 
\end{align}
and
\begin{align}\label{eq:fGW}
f_{\rm GW} &= 6.3 \times 10^{-9}
\left(\frac{T}{10~{\rm year}}\right)^{-1}
~{\rm Hz}  ,
\end{align}
where 
$f_{\rm GW}=2/T$, and
${\cal M_{\rm ch}}=
\left(\frac{M_{1}M_{2}}{M}\right)^{3/5}M^{2/5}$,
are
the frequency of the emitted GWs, and 
the chirp mass of the binary, respectively
\citep[][]{schutz16}.
The chirp mass is known to be rewritten as
${\cal M_{\rm ch}}
=M_{1}\frac{q^{3/5}}{(1+q)^{1/5}}$.
Since $q\le1$, the chirp mass follows
${\cal M}_{\rm ch}=\frac{q^{3/5}}{1+q^{6/5}}M\le 2^{-6/5}M \approx 0.435M$.
The ${\cal M}_{\rm ch}$ is at a maximum for $q=1$ and decreases monotonically as $q$ decreases.
The GWB strain amplitude constrained by the NANOGrav15  roughly indicates $h_{\rm GWB} \approx 1\times 10^{-14}$ \citep[see Figure~1 in][]{agazie23b}.
To avoid the overproduction of the GWB as suggested by the NANOGrav15 dataset, the condition $h_{\rm GWB} \ge h_{c}$ is satisfied.
Then, one can obtain the lower limit $a_{\rm GWB}$ 
in the frequency range of $2~{\rm nHz} \lesssim f_{\rm GW} \lesssim 30~{\rm nHz}$
as follows:
\begin{align}\label{eq:a_GWB}
a_{\rm GWB} = 
1.4\times 10^{17}
\left( \frac{M_{1}}{10^{9}M_{\odot}}  \right)^{2}
\left(\frac{D}{10~{\rm Mpc}}\right)^{-1}
\left(\frac{h_{\rm GWB}}{1\times 10^{-14}}\right)^{-1}
q   ~{\rm cm},
\end{align}
where the relations of 
$f_{\rm GW}=(GM/a_{\rm GWB}^{3})^{1/2}/\pi$ and 
${\cal M}_{\rm ch}=\frac{q^{3/5}}{1+q^{1/5}}M_{1}$
are used to obtain Eq.~(\ref{eq:a_GWB}).
Thus, the NANOGrav 15-year dataset can place an upper limit
through the GWB strain amplitude.

\subsection{Persistent EHT ring images}

As $a$ decreases to sub-parsec scale,
we must account for $t_{\rm GW}\propto a^{4}$, 
as $t_{\rm GW}$ becomes significantly short.
If M87 were in a merger phase, the black hole shadow and its surrounding ring image of M87 would exhibit highly nonlinear and dynamic features
\citep[][]{yumoto12,bohn15,cunha18}.
However, EHT observations clearly show the
persistent ring image, at least in 2017 and 2018 \citep{EHT19_1,EHT24}, allowing us to
readily exclude that the binary system
is already in a merging phase.

When $t_{\rm GW}$ is shorter than the binary orbital period ($T$),
the binary system
is expected to merge into a single SMBH
(see  Figure~\ref{fig:timescale}).
Based on this criterion, 
we define the timescale $t_{\rm merge}$,
which satisfies
$t_{\rm merge} \equiv  t_{\rm GW}(a=a_{\rm merge})$.
Using this definition, 
we find that 
$t_{\rm merge}\approx 0.3$~years ($q=1$) 
at $a\approx 1.3\times10^{16}$~cm
for M87 (see Figure~\ref{fig:timescale}).
The corresponding lower limit $a$
below which the 
binary black holes are expected to merge
into a single SMBH
can be defined by
\begin{align}\label{eq:a_merge}
a_{\rm merge} = 
\left[\frac{256G^{3}M_{1}^{3}t_{\rm merge}}{5c^{5}}  q(1+q) \right]^{1/4} .
\end{align}
Due to the weak dependence of  $a_{\rm merge}$
on $t_{\rm merge}$, i.e., 
$a_{\rm merge}\propto t_{\rm merge}^{1/4}$,
the uncertainty in $t_{\rm merge}$ does not significantly impact the estimate of 
$a_{\rm merge}$.

\section{Results}\label{sec:result}

Here, we present the results of applying the SMBH binary model to M87.
As illustrated in Figure~\ref{fig:geometry}, 
we consider the situation where
$M_{1}$ generates the observed prominent radio jet in M87, while $M_{2}$ does not produce a jet.


\subsection{Comparison of Timescales}

\begin{figure} 
\includegraphics
[width=18cm]
{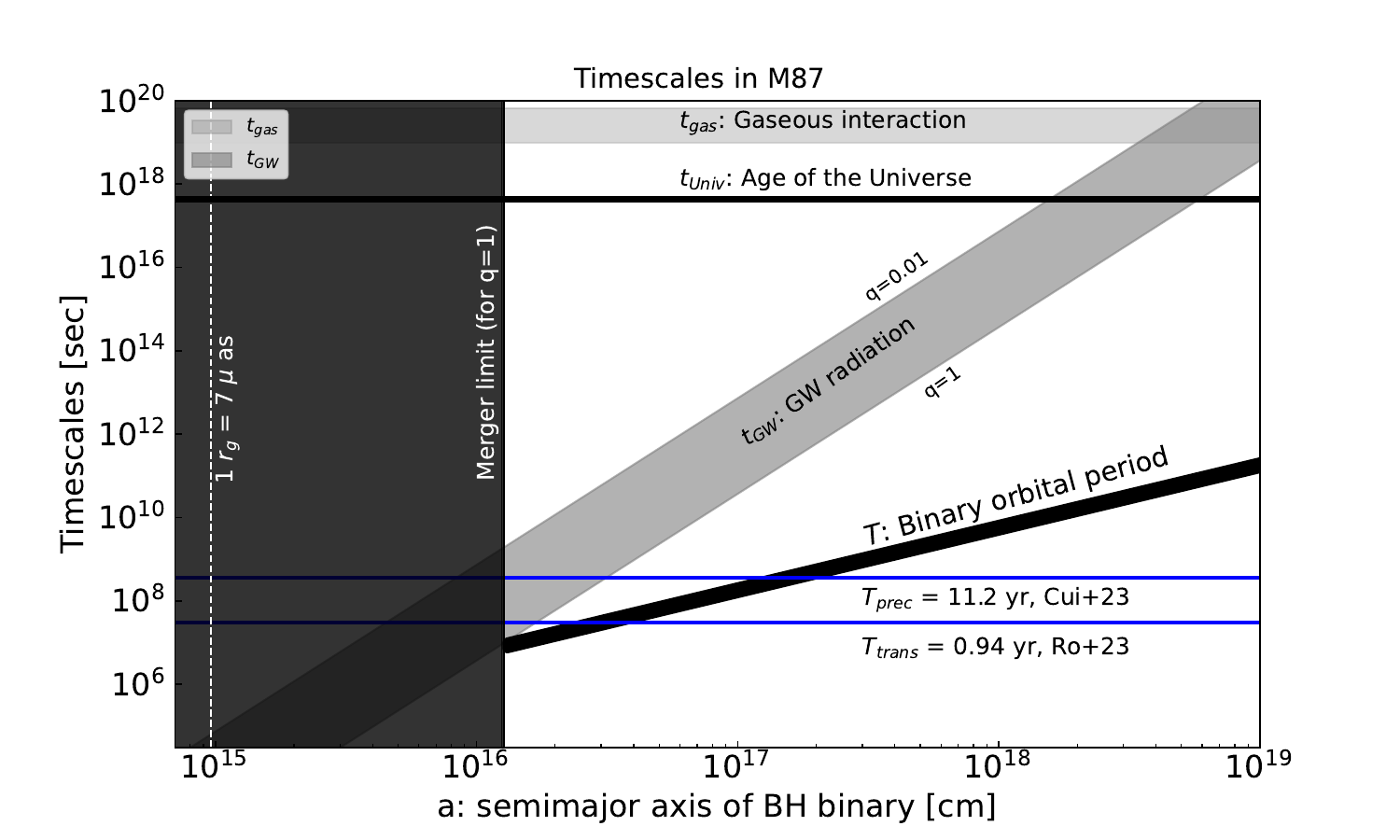}
\caption
{Comparison of characteristic timescales in M87.
The binary orbital period ($T\propto a^{3/2}$), represented by the thick line, 
is the most important timescale, 
as given by Eq.~(\ref{eq:T}).
The timescale for GW radiation, $t_{\rm GW}\propto q^{-1} a^{4}$, is represented by a dark-gray thick line for $0.01\le q\le 1$.
We also plot the two blue lines representing $T_{\rm prec}$ and $T_{\rm trans}$.
When $T=T_{\rm prec}$, the corresponding semimajor axis is approximately
$a\approx 1\times 10^{17}~{\rm cm}$.
When $T=T_{\rm trans}$, the corresponding semimajor axis is approximately 
$a\approx 3\times 10^{16}~{\rm cm}$.
When $t_{\rm GW}<T$, the SMBH binary will merge into a single SMBH. This occurs when semimajor axis is less than approximately 
$a \lesssim 1\times 10^{16}~{\rm cm}$.
In addition, $t_{\rm gas}$ for M87 is shown in light gray, and is even longer than $t_{\rm Univ}$ due to its small mass accretion rate of $\dot{m}\sim 10^{-6}$.
}
\label{fig:timescale}
\end{figure}

Figure~\ref{fig:timescale} presents a comparison of relevant timescales in M87.
The orbital period ($T$) and 
the timescale of GW radiation ($t_{\rm GW}$) 
are given by Eqs~(\ref{eq:T}) and (\ref{eq:t_GW}),
respectively.
To plot $T$ and $t_{\rm GW}$, 
we set $0.01\le q \le 1$.
For plotting $t_{\rm gas}$, we set $q=0.01$ and 
choose the mass accretion rate as
$\dot{M} = (3-20) \times 10^{-4}~M_{\odot}~{\rm yr^{-1}}$, based on the result of \citet{EHTC8}.
Figure~\ref{fig:timescale} shows
that the range of semimajor axis  
where  $T$ is comparable to $T_{\rm prec}$ or $T_{\rm trans}$ lies within approximately
$ 10^{16}~{\rm cm} \lesssim a \lesssim 10^{17}~{\rm cm}$, 
well-aligned
with spatial resolutions of 
recent VLBI observations. 
This makes M87 the best candidate for starting a detailed study of a potential SMBH binary system.
It is well known that
a criterion that GWs become important for binary evolution can be estimated as 
$t_{\rm GW} \le t_{\rm Univ}\approx 1.38\times 10^{10}$~yr \citep[][for review]{dorazio23}.
From Figure~\ref{fig:timescale},
one can see that this condition is met when $a\lesssim 10^{18}$~cm for M87.
\footnote{A pioneering study by \citet{yonemaru16} 
investigated GW emission from M87. However, their analysis considered a larger 
$a$ compared to the present work.}

Compared to $t_{\rm gas}$ discussed in this work and previous studies \citep{dorazio18,zhao24}, 
a notable difference is identified, i.e.,  $t_{\rm gas}$ appears excessively long in the case of M87.  
This is naturally understood, as both of the aforementioned prior studies \citep{dorazio18,zhao24} considered a mass accretion rate of $\dot{m}\sim 1$, whereas M87, on the contrary, has the mass accretion rate on the order of $\dot{m}\sim 10^{-6}$, which is substantially smaller than $\dot{m}\sim 1$.
Thus, at first glance, gaseous interactions appear
negligible in the current state of M87. 
However, $t_{\rm gas}$ exceeds the age of the Universe $t_{\rm Univ}\approx 1.38\times 10^{10}$~yr, raising well-known "final parsec problem", -- namely, how a binary system can form under such circumstance \citep[e.g.,][for review]{milosav03,dorazio23}.
One possibility may be the presence of a specific period during which $\dot{m}$ becomes significantly elevated. Interactions and mergers between galaxies are known to trigger large-scale nuclear gas inflows, supplying gas to SMBHs \citep[e.g.,][]{hernquist89, diMatteo05, hopkins06, hopkins08}. Such galaxy interaction or merger events could promote temporarily elevated $\dot{m}$ levels.
Another possibility is the presence of cold gas. 
Recent ALMA CO(1–0) observations reported by \citet[][]{ray24} reveal the distribution of molecular clouds within approximately 100 parsecs of the M87 nucleus. This newly identified cold gas could potentially reduce $t_{\rm gas}$.
Additionally, unequal-mass SMBH binaries are known to show significantly higher eccentricity 
than equal-mass SMBH binaries, 
which may accelerate the decrease in $a$ \citep[e.g.,][]{mikkola92, matsubayashi07, enoki07, iwasawa11}.
Since the primary aim of this paper is not to investigate the formation of a binary system in M87 in detail, but rather to constrain the allowed range of $q$ based on the latest observational constraints, we do not undertake detailed modeling of $t_{\rm gas}$ in M87.

\subsection{Allowed parameter space for $q$ and $a$}

In Figure~\ref{fig:q-a}, we present the allowed 
parameter space for $q$ and $a$ in M87.
All the gray-shaded regions represent the excluded areas for $q$ and $a$, while the remaining white region is the allowed parameter space for $q$ and $a$.
As we consider the SMBH binary system, the upper limit of $a$ is inherently constrained by the condition of $a\geq a_{\rm hard} \approx 5 \times 10^{19} \left(M_{1}/10^{9}M_{\odot}\right){\rm cm}$, assuming $\sigma_{\star} \approx 300~{\rm km~s^{-1}}$.

\subsubsection{The Cases of $T=T_{\rm prec}$ and $T=T_{\rm trans}$}

\begin{figure} 
\includegraphics
[width=18cm]
{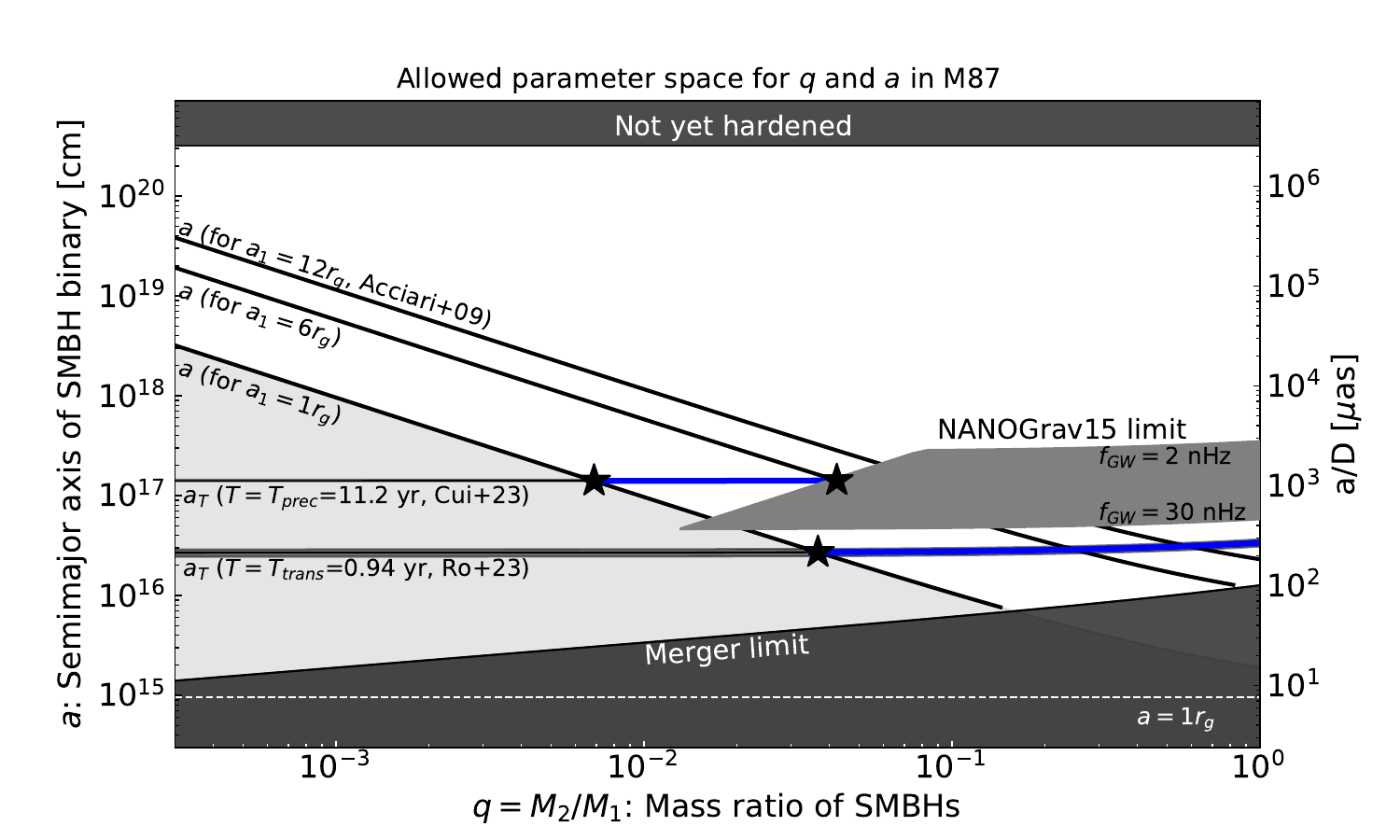}
\caption
{Allowed parameter space for $q$ and $a$ in M87.
All the gray-shaded regions represent the excluded areas for $q$ and $a$, while the remaining white region is the allowed parameter space for $q$ and $a$.
In particular, the allowed ranges for the Cases of $T=T_{\rm prec}$ are $T=T_{\rm trans}$ are shown in blue lines.
The upper limit of $a$ is constrained by the condition of $a < a_{\rm hard}$.
The lower limit of $a$ is constrained by the condition of $a > a_{\rm merge}$.
The upper limit of $q$ is partially constrained by the GWB strain amplitude obtained by the NANOGrav15 \citep{agazie23a}. 
This constraint applies only to the range of 
$2~{\rm nHz} \lesssim f_{\rm GW} \lesssim 30~{\rm nHz}$.
The lower limit of $a_{1}$ is determined by the existence of a reflex motion of $M_{1}$, shown as the light gray-shaded region, where $a_{1} > 1~r_{g}$ is assumed.
In the cases of $T=T_{\rm prec}$ and  $T=T_{\rm trans}$, the lower limit of $q$ is determined 
by $T(a_{1}=1~r_{g}) = T_{\rm prec}$ or $T(a_{1}=1~r_{g}) = T_{\rm trans}$ and is marked as stars. 
The upper limit of $q$ is bounded by the
NANOGrav 15 limit in the case of $T=T_{\rm prec}$.
}
\label{fig:q-a}
\end{figure}

Here, we consider the Cases of 
$T=T_{\rm prec}$ and $T=T_{\rm trans}$.
To constrain $q$ and $a$, we impose
the following three constraints:

\begin{enumerate}

\item 
The lower limit of $a$ is constrained by the condition of $a > a_{\rm merge}$.
When the semimajor axis contracts to $a\le a_{\rm merge}$ as given by Eq.~(\ref{eq:a_merge}), the SMBH binary undergoes a merger due to the loss of angular momentum and energy through GW radiation. 
The black region below the "merger limit" corresponds to this case.
The entire range for $T = T_{\rm prec}$ or $T = T_{\rm trans}$ remains unaffected by this "merger limit".

\item
The region excluded by the NANOGrav15 observation is shaded in dark gray.
To avoid the overproduction of the GWB as suggested by the NANOGrav15 dataset, the condition $a \ge a_{\rm GWB}$ is imposed at the frequency range 
$2~{\rm nHz} \lesssim f_{\rm GW} \lesssim 30~{\rm nHz}$.
As mentioned in sub-section \ref{sec:GWB}, 
here we set $h_{\rm GWB} =1\times 10^{-14}$ based on \citet{agazie23a}.
From Figure~\ref{fig:q-a}, we find that
the upper limit of the allowed range
is bounded by the NANOGrav 15-year limit in the case of $T=T_{\rm prec}=11.2~{\rm yr}$.
In contrast, the case of $T=T_{\rm trans}=0.94~{\rm yr}$ is not excluded by the NANOGrav 15-year limit
because
the frequency range covered by the NANOGrav does not extend beyond $ f_{\rm GW} \approx 30~{\rm nHz}$.

\item
A lower limit of $a_{1}$ is set here.
As illustrated in Figure~\ref{fig:geometry}, the reflex motion of $M_{1}$ induces periodic behavior in the jet for both $T = T_{\rm prec}$ and $T = T_{\rm trans}$. Consequently, $a_{1}$ cannot be zero and must have a finite value.
Since the jet behavior at different $a_{1}$ values is not well investigated, we assume a lower limit of $a_{1} = 1~r_{g}$, labeled as $a$ (for $a_{1} = r_{g}$). Below this threshold, the jet is unlikely to exhibit a distinct periodic signature.

\end{enumerate}

By combining all the constraints described above, we obtain the allowed range of $q$ as follows:
\begin{align}
& 6.9\times 10^{-3} \le   q \le  4.2 \times 10^{-2}
\quad ({\rm for} ~T=T_{\rm prec}=11.2~{\rm yr}), \\
& 3.7 \times 10^{-2} \le q \le 1  
\quad ({\rm for} ~ T=T_{\rm trans}=0.94~{\rm yr}).
\end{align}
The blue lines in Figure~\ref{fig:q-a} represent the allowed ranges for these cases.
The intersection points of the excluded regions and and $a_{T}$ are marked by the star-shape dot to facilitate visual identification of the allowed parameter ranges.

\subsubsection{
The Case of neither $T=T_{\rm prec}$ nor $T=T_{\rm trans}$}

Next, we consider the Case of neither $T=T_{\rm prec}$ nor $T=T_{\rm trans}$.
In this case, the entire white region
 in Figure~\ref{fig:q-a} is the allowed parameter space for $q$ and $a$, resulting in a significant expansion of the allowed parameter space.
The origin of the observed jet precession \citep{cui23} and the transverse oscillation \citep{ro23_trans} 
should not be a reflex motion in the SMBH binary.
Instead, it should be naturally attributed as suggested in each respective paper.

Although merger limit and the NANOGrav15 limit offer important limits
on the allowed parameter space, it is intriguing to identify a gap-window region in between them where the case $q\sim 1$ remains valid.
Coincidentally, the gap-window region envelopes the aforementioned case of $T=T_{\rm trans}$.
Similar to the case  mentioned above,
the gap-window region with $q\sim 1$ aligns with the case of approximately $a_{1}\sim 10~r_{g}$.
Proposed approaches for investigating  the allowed parameter space will be also discussed in Section~\ref{sec:discussion}.

Additionally, it is worth noting that a previous work by \citet[][]{safarzadeh19} also addressed $q$ in M87.
However, a key difference between their study and the present work lies in the GW frequency range to which the PTA limit is applied.
The upper limit of $a_{\rm T}$ approximately 
0.01~pc in the PTA limit adopted in \citet[][]{safarzadeh19} (the black hatching area in Figure~1 of their paper) corresponds 
to the GW frequency range of $f_{\rm GW} \gtrsim 50$~nHz, based on Eqs.~(\ref{eq:a_T}) and (\ref{eq:fGW}).
However, the effective frequency range of NanoGrav 15-year dataset is $2~{\rm nHz} \lesssim f_{\rm GW} \lesssim 30~{\rm nHz}$ \citep{agazie23a}. 
This difference in the GW frequency range 
likely explains why the gap-window region found in this work does not appear in \citet[][]{safarzadeh19}.

\section{Discussion}\label{sec:discussion}

\subsection{Astrometric observations  with
current VLBI facilities}

It is widely recognized that VLBI astrometric observations, through the direct tracking of orbital motions, could provide conclusive evidence of sub-parsec separation SMBH binaries on orbital timescales \citep[e.g.,][for review]{dorazio23}.
\footnote{ Astrometric monitoring observations in near-infrared band,
aimed at determining SMBH binaries, are also discussed by \citet{dexter20}.}
However, one significant limitation of VLBI astrometry is the requirement for expensive, multi-epoch VLBI observations, restricting its feasibility to a limited number of targets.
Therefore, we initiate an investigation of M87 as a likely target system, given the periodic behavior observed in its jet.
\footnote{
In contrast to M87, 
OJ287 ($z=0.306$) is a well-discussed 
SMBH binary candidate due to its
repeated double-peaked outburst features in the optical band with ~12-year intervals and 
complex transverse motion of its jet
\citep[e.g.,][]{Sillanpaa88, britzen18,britzen23}. 
Recently, \citet[][]{cheng23} discussed 
a feasibility for future VLBI astrometry, emphasizing the importance of a suitable reference source.}
Below, we briefly discuss strategies for conducting VLBI astrometric observations using current VLBI facilities.

Our ultimate science goal is to identify definitive evidence for the existence of close separation SMBH binaries nearing mergers. 
Therefore, cases with larger $q$ would be more intriguing than those with smaller $q$.
In this context, the gap-window region between the NANOGrav15 limit and the merger limit 
appeared in Figure~\ref{fig:q-a} 
is of great interest for investigation through VLBI astrometric observations.
The upper limit of the gap-window region is defined by the upper bound of the NANOGrav's frequency range, $f_{\rm GW}=30$~nHz, which corresponds to a binary orbital period of $T\approx 2$~years.
To investigate the presence of reflex motion in the gap-window region, it is crucial to detect at least one full cycle of periodic motion or a slightly longer duration. This corresponds to approximately $2-3$ years.
As a first step, it would be reasonable to begin exploration in the gap-window region around approximately $\sim 40~\mu$as ($a_{1}\approx 6~r_{g}$), comparable to previous works, rather than attempting to address the scale of
 $\sim 4~\mu$as ($a_{1}\approx 1~r_{g}$) from the beginning.
The position error ($\Delta \theta$) originated 
from the phase error ($\Delta\phi$) is generally expressed as follows:
\begin{align}
\Delta \theta = 
\frac{\lambda_{\rm obs} }{2\pi D_{\rm bl} }\Delta\phi
\approx c \Delta\tau_{0} \sec Z \tan Z \Delta Z
\end{align}
where 
$D_{\rm bl}$, 
$\lambda_{\rm obs}$,
$\Delta\tau_{0}$, and
$Z$,
are 
the baseline length, 
the observing wavelength, 
the residual vertical delay, and
the local source zenith angle,
respectively 
\citep{thompson01,reid99}.
For instance, it can be estimated as 
$\Delta \theta \sim 28~{\rm \mu as}$ 
at 43~GHz (7~mm)
for the tropospheric zenith delays within 
$\Delta \tau_{0} \approx 2$ cm accuracy
for VERA array \citep[][]{honma08}, 
$\Delta Z$ = 0.5 deg, 
Z = 50 deg and $D_{\rm bl} \approx 2\times 10^{3}$~km
\citep[e.g.,][]{koyama15,niinuma15}.
While VLBI astrometry observations can directly constrain $a_{1}$, only a limited number of such observations have been conducted for M87 in the past \citep{acciari09, hada12,hada14}.
Figures \ref{fig:timescale} and \ref{fig:q-a} 
clearly indicate that VLBI astrometric monitoring
of M87's jet base over just a few months is insufficient. 
Long-term monitoring would be essential to better constrain the allowed ranges for $a$ and $q$.
Given the practical challenges of conducting continuous VLBI astrometry observations over a 10-year period, we propose a more feasible approach: initiating a 1–2 year VLBI astrometry pilot study of M87 to obtain initial results.
If the pilot data indicate a systematic motion of $a_{1}$, it would motivate the continuation of longer-term observations.

One potential caveat would be 
overlapping of newly ejected
blob component(s) onto the underlying continuous
jet image during flaring events.
\citet{hada14} conducted VERA astrometry
for the M87 core with respect to the core position of M84, during the VHE flaring event in 2012.
They detected the core shifts between 22 and 43 GHz,
with the mean value of ($\Delta x_{22-43}$,$\Delta y_{22-43}$)=(64, 95)~$\mu$as. 
The radio core flux densities showed frequency-dependent evolution, with more rapid increases at higher frequencies and greater amplitude variations.
The light curves revealed a time lag between the peaks at 22 and 43~GHz, constrained to approximately 35–124~days. This suggests that a newly born radio-emitting component was generated near the black hole during the VHE flaring event in 2012 and subsequently propagated outward with the speed of 
$\sim (0.04 - 0.22)~c$.
Such a propagation of the new component may introduce errors in estimating the location of $M_{1}$. However, since the duration of the VHE flaring event is limited to less than a month \citep{hada24}, and the newly born radio-emitting component can be identified through VLBI observations \citep[][]{hada14}, it may be possible to minimize 
errors in estimating the location of $M_{1}$ caused by the flare-associated component.

Another caveat could be an annual parallax
when exploring $T\approx 1~{\rm year}$.
Annual parallax is geodetic effects related to Earth’s orbital motion around the Sun. It is based on the principle of triangulation. This shift is caused by the change in perspective as Earth moves from one side of its orbit to the other, creating a slight change in the angle from which we view 
a target source.
Therefore, if $T=T_{\rm trans}\approx 1~{\rm year}$ is the case, then one have to carefully discriminate the reflex motion to the annual parallax \citep[][]{sudou03}.
With the definition 
$1~{\rm arcsecond} \equiv 1~{\rm au}/1~{\rm pc}$,
the expected amplitude of the annual parallax
for $D=16.8$~Mpc corresponds to $0.06~{\mu}$as,
which is significantly 
smaller than the target accuracy of this work.
Moreover, long period observations will
help overcome challenges in
exploring $T\approx 1~{\rm year}$.
If astrometric observations are continuously conducted over a long period of 10~years
assuming this is practically feasible,
the binary is expected to complete 9.4 orbits,
which should be distinguishable from 
completing 10~orbits.

Multi-frequency receivers designed for VLBI astrometry are currently being installed or scheduled to be installed on many VLBI telescopes worldwide \citep[][for review]{dodson23}.
In the near future,
multi-frequency VLBI systems will significantly advance our ability to address key scientific questions.
In particular,
phase-referencing observations at 86~GHz, utilizing the frequency phase transfer technique \citep[][]{rioja20,dodson23,issaoun23}, will enable us to perform phase-referencing observations illustrated in Figure~\ref{fig:geometry}.
This is because 86~GHz imaging can reveal the ring-like accretion structure image at the M87 jet base \citep[][]{lu23,kim24}, facilitating
for more accurate tracking of 
$M_{1}$, which is likely located within this structure.
However, a major confounding effect in extracting potential orbital motion is expected to arise from structural variations in the jet base region, such as an ejection of a new jet component and/or jet-disk interactions.
Future projects of 
the next-generation Event Horizon Telescope (ngEHT) \citep[][]{doeleman23,johnson23ngEHTsci} 
and the Black Hole Explorer (BHEX) \citep[][]{johnson24},
will be capable of capturing detailed structural changes in the jet base region.
Since such structural variations may complicate the extraction of pure orbital motion of $M_{1}$, ngEHT and BHEX will play a crucial role in mitigating the confounding effect.

\subsection{Future astrometric observations with ngVLA} 

\citet{wrobel22} highlighted that the next-generation Very Large Array \citep[ngVLA;][]{murphy18} will be 
a powerful tool 
for astrometric observations of SMBH binary candidates. 
They noted that using multiple phase calibrators, separated from a target source by less than 1–2 degrees for the phase-referencing observation, would allow achieving position accuracy levels on the order of $1~\mu$as at millimeter wavelengths
\citep[see also][]{rioja20}.
From Figure~\ref{fig:q-a}, it is clear that position accuracy levels on the order of $1~\mu$as correspond to the region below the "merger limit" in the case of M87. 
Therefore, ngVLA astrometric observations may be unnecessarily precise for M87 when searching for a hypothetical secondary black hole.
However, it is worth identifying intriguing scientific cases with future ngVLA  observations.

We emphasize that
the below-merger-limit domain corresponds to the case where M87 has a single black hole. Therefore, this domain offers a valuable opportunity to avoid the potential overlap of multiple origins for any positional shifts of $M_{1}$ that future ngVLA astrometric observations might detect.
In this context, we discuss the feasibility of searching for fuzzy dark matter (FDM) at the center of M87. A SMBH at the center of a star cluster or galaxy experiences Brownian motion due to gravitational encounters with stars, leading to displacement from its central position \citep[e.g.,][]{bahcall76, merritt07}. 
The nuclear point (SMBH location) of M87 appears to be offset from the galaxy’s photo-center by about 6~pc \citep[][]{batcheldor10}. This off-center displacement can be attributed to gravitational interactions with stars over 10 billion years \citep{DiCintio20}.
The expected speed of this gravitational Brownian motion due to stars $\sigma_{\rm BH/\star}$
can be approximately estimated as
\begin{align}\label{eq:sigma_BHstar}
    \sigma_{\rm BH/\star}\approx 
    \left(\frac{m_{\star}}{M}\right)^{1/2} 
    \sigma_{\star}
    \approx 0.001~{\rm km~s^{-1}}
        \left(
        \frac{\sigma_{\star}}{100~{\rm km~s^{-1}}} 
        \right)  ,
\end{align}
where 
$m_{\star}$ and $\sigma_{\star}$ are
the mass of a typical star, and
the stellar velocity dispersion, respectively
\citep{reid04,merritt07}
\footnote{Originally, \citet{reid04,merritt07} denoted the speed of this gravitational Brownian motion as $<V>$. Here, we denote it as $\sigma_{\rm BH/\star}$ for convenience.}
and Eq.~(\ref{eq:sigma_BHstar}) 
is the case for  $m_{\star}/M\approx 10^{-10}$.
However, gravitational Brownian motion is not only caused by stars but also by the surrounding dark matter. It generally reflects the physical properties of the entire surrounding environment.
One of the interesting explorations for the
surrounding environment could be constraining the mass of dark matter via gravitational Browninan motion of SMBHs in a dark matter halo.
FDM, consisting of ultralight axions, has been proposed to mitigate galactic-scale problems within the cold dark matter scenario
\citep[e.g.,][for review]{ferreira21}.
\citet{elzant20}
assume that the SMBH achieves equilibrium with the fluctuations, that is, there is a balance between the effects of fluctuation and dissipation, the latter being due to dynamical friction. 
\footnote{
\citet{kawai22,kawai24} suggest, on the contrary, that  FDM is likely to have a smooth density distribution without such heavy
FDM quasi particles in the central region of an FDM halo. If this is the case, then the argument
in this sub-section may no longer hold.}
Assuming the energy equi-partition condition, the velocity dispersion of the SMBH caused by the FDM
($\sigma_{\rm BH/FDM}$) is related 
to the FDM velocity dispersion ($\sigma_{\rm halo}$) as
$M\sigma_{\rm BH/FDM}^{2}
=m_{\rm eff}\sigma_{\rm halo}^{2}/2$
where $m_{\rm eff}$
denotes the effective mass of FDM quasi-particles.
Then, the velocity dispersion of the SMBH caused by the FDM is given by
\begin{align}
    \sigma_{\rm BH/FDM}
    \approx \left(\frac{m_{\rm eff}}{M}\right)^{1/2}     
    \sigma_{\rm halo}
    \approx 
25~{\rm km~s^{-1}}
\left(\frac{m_{\rm ax}c^{2}}{10^{-22}~{\rm eV}}\right)^{-1/2}
\left(\frac{M}{6\times 10^{9}~{M_\odot}}\right)^{-1/6}
 f(x)^{1/2}   ,
\end{align}
where 
$f(x)=(1+x^{2}/3)(x^{2}+3)^{2}/[(1+x^{2})(x^{2}+2)]^{2}$,
$x=r/r_{\rm core}$, and 
$m_{\rm ax}$ is the axion mass
\citep[see details][]{elzant20}.
Interestingly,
$\sigma_{\rm BH/FDM} \gg \sigma_{\rm BH/\star}$
is expected.
The corresponding angular velocity of SMBH Brownian motion, however, is less than $1~{\rm \mu as~yr^{-1}}$.
Another difficulty may be 
an existence of secular proper motion.
As pointed out by \citet{wrobel22},
there could be relative motion even between M87 and a reference source. 
M87 is located at the center of the Virgo cluster, and reference sources commonly used for astrometric observations of M87 (e.g., M84) are also sufficiently within the virial radius of the Virgo cluster
\citep[about $10^{3}$ kpc][]{simionescu17} and reach virial equilibrium.
Recently, the infalling velocity M49 group located outside the virial radius of the Virgo cluster, has been estimated as 
$v_{\rm infall}\approx 300-640~{\rm km~s^{-1}}$ by \citet{su19}, corresponding to
$\dot{\theta}_{\rm infall}\approx 4-8~{\rm \mu as~yr^{-1}}$.
The $\dot{\theta}_{\rm infall}$ would be superposed onto the possible gravitational Brownian motion 
of SMBH by FDM.
Therefore, detecting $\sigma_{\rm BH/FDM}$ may not be straightforward.
To achieve high-precision astrometry with $\dot{\theta} < 1~{\rm \mu as~yr^{-1}}$, astrometric observations at 86 GHz or higher would be required \citep[e.g.,][]{jiang23, issaoun23, zhao24}.

\subsection{Towards localization of GWB sources}

\citet{HD83} highlighted that 
the angular correlation pattern
for pulsar pairs averaged over the whole sky domain 
can serve an evidence of GWs, as the resulting correlation curve reflects a quadrupole nature of GWs \citep[e.g.,][]{maggiore18}.
However, 
the whole-sky average makes the Hellings-Downs curve insensitive to any particular direction of the sky.
To overcome this problem, \citet{sasaki24} explored
what happens in the pulsar correlation if the averaging domain is changed from the whole sky.
They found that the angular correlation pattern for pulsar pairs within a chosen sky Hemisphere has a dependence on a single GW compact source.
They indicated that if a single GW source is dominant, the variation in a Hemisphere-averaged angular correlation curve is 
the greatest when the chosen Hemisphere has its North Pole at the sky location of the GW source.
According to \citet{sasaki24}, a nearby GW source can be marginally detected when the distance ($D$) to the  GW source is
\begin{align}
    D \lesssim 2\times 10^{2} ~{\rm Mpc}
    \left(\frac{N_{\rm pulsar}}{10^{2}} \right)^{1/2}
    \left(\frac{\Delta \Gamma_{H}}{0.1} \right)^{1/2}
         \left(\frac{\Delta t_{a}}{1~{\rm \mu~sec}} \right)
     \left(\frac{M}{6.5\times 10^{9}~M_{\odot}} \right)^{5/3}
         \left(\frac{f_{\rm GW}}{1~{\rm yr^{-1}}} \right)^{1/3}  ,
\end{align}
where
$N_{\rm pulsar}$,
$\Gamma_{H}$,
$\Delta \Gamma_{H}$, and 
$\Delta t_{a}$ are
the total number of the observed pulsars in PTAs, 
the difference between the maximum and minimum of
the Hemisphere-averaged cross-correlation of pulsar pairs
($\Gamma_{H}$),
the variation of $\Gamma_{H}$ 
for changing the inclination angle of the Hemisphere from the GW source direction, and 
the delay in the measured pulse arrival time from the 
the $a$-th pulsar relative to the expected arrival time in the absence of gravitational waves,
respectively.
From this, it is clear that the distance of M87,
 $D=16.8$~Mpc, falls well
within the target range of this Hemisphere-averaged method.

Enhancing PTA sensitivities is an important factor
on searching for GWB sources.
There are widely-discussed two ways to realize it.
The first one is constructing Square Kilometer Array (SKA).
The SKA would significantly improve the sensitivity of PTAs through newly finding hundreds of millisecond pulsars (MSPs) 
(https://www.skao.int/en). 
Then, SKA will surely facilitate to perform 
sky localization by using the Hemisphere-averaged correlations.
The other one is suppression of uncertainties of MSPs.
One of the limiting factors on searching for the GWB with PTAs is the uncertainties on the Solar system planetary ephemerides \citep[e.g.,][]{Vallisneri20}, 
which are used to convert geocentric time-of-arrivals of pulses to ones measured in the Solar system barycentric frame.
VLBI astrometry of MSPs can suppress those uncertainties, 
and hence enhance the PTA sensitivities 
\citep[][and references therein]{smits11, siemens13, ding23,kato23}.

\section{Summary}

In this paper, 
we constrain a mass of a hypothetical secondary black hole orbiting the primary SMBH in M87.
To constrain $q$ and $a$, 
we impose the following three constraints:
(i) the lower limit of $a$, below which the 
SMBH binary is expected to merge.
(ii) the strain amplitude of the GWB 
shown in the NANOGrav 15-year dataset.
(iii) 
a finite $a_{1}$ that can induce periodic behavior in the jet.
By combining these constraints, we obtain the allowed parameter space for $q$ and $a$.
If either of the EAVN-detected periods ($T$) corresponds to the binary's orbital period, 
the allowed range of $q$ is $6.9 \times 10^{-3} \le q \le 
4.2 \times 10^{-2}$ for $T \approx 11$ years, and 
$3.7 \times 10^{-2} \le q \le 1$
for $T \approx 0.9$ years.
VLBI astrometric monitoring of the jet base of M87 is essential to explore the allowed parameter space for $q$ and $a$.

VLBI astrometric observations enables detection of sub-parsec separation SMBH binaries by tracking orbital motions, but their high cost and long duration limit target numbers. 
M87, with its jet periodicity, is a good candidate, particularly in the gap-window region between the NANOGrav limit and the merger limit. A 1–2 year pilot study could reveal systematic motion, prompting longer-term observations. 
Challenges such as jet flares and annual parallax may be mitigated by multi-frequency phase-referencing VLBI techniques. Upcoming systems, especially at 86 GHz, will improve positional accuracy and help test the SMBH binary hypothesis.

We discuss the future potential of ngVLA astrometry, expected to achieve $\sim 1~\mu$as accuracy. While exceeding the requirements for detecting a secondary black hole in M87, this precision could investigate gravitational Brownian motion from ultralight FDM. 
Detecting such motion is challenging due to M87's possible secular motion in the Virgo cluster and the required sub-$1~\mu$as~yr$^{-1}$ precision, while future instruments like ngEHT and space VLBI may improve accuracy and uncover insights into FDM.

\bigskip
\leftline{\bf \large Acknowledgment}
\medskip

\noindent

We thank the anonymous referee for providing valuable
comments, which improved the paper.
This work was partially supported by
the MEXT/JSPS KAKENHI
(JP21H01137,
JP21H04488,
JP22H00157,
JP23H00117, and
JP23K03448).
H.R. is supported by the National Research Council of Science \& Technology (NST) -- Korea Astronomy and Space Science Institute (KASI) Postdoctoral Fellowship Program for Young Scientists at KASI in South Korea.
Y.C. is supported by the Natural Science Foundation of China (grant 12303021) and the China Postdoctoral Science Foundation (No. 2024T170845).
This work was supported by the National Research Foundation of Korea (NRF) grant funded by the Korea government (MSIT; RS-2024-00449206) and the POSCO Science Fellowship of POSCO TJ Park Foundation.

\newpage

\bibliography{ms.bbl}{}




\end{document}